\documentclass[a4paper]{jpconf}
\usepackage{graphicx}
\usepackage[usenames,dvipsnames]{xcolor}
\begin{document}
\title{Electron irradiation of Co, Ni, and P-doped BaFe$_{2}$As$_{2}$--type iron-based superconductors}

\author{Cornelis Jacominus  van der Beek$^{1}$, Sultan Demirdi\c{s}$^{1}$,  Doroth\'{e}e Colson$^{2}$, Florence Rullier-Albenque$^{2}$, Yanina Fasano$^{3}$, Takasada Shibauchi$^{4}$, Yuji Matsuda$^{4}$, Shigeru Kasahara$^{4,5}$, 
Piotr Gierlowski$^{6}$, Marcin Konczykowski$^{1}$}
\address{$^{1}$Laboratoire des Solides Irradi\'{e}s, CNRS UMR 7642 \& CEA-DSM-IRAMIS, Ecole Polytechnique, F-91128 Palaiseau cedex, France}
\address{$^{2}$Service de Physique de l'Etat Condens\'{e}, CEA Saclay, CEA-DSM-IRAMIS, CNRS URA 2464, F-91191 Gif-sur-Yvette, France}
\address{$^{3}$Laboratorio de Bajas Temperaturas, Centro At\'{o}mico Bariloche \& Instituto Balseiro, Avenida Bustillo 9500, 8400 San Carlos de Bariloche, Argentina}
\address{$^{4}$Department of Physics, Kyoto University, Sakyo-ku, Kyoto 606-8501, Japan}
\address{$^{5}$Research Center for Low Temperature and Materials Sciences, Kyoto University,Sakyo-ku, Kyoto 606-8501, Japan}
\address{$^{6}$Institute of Physics of the Polish Academy of Sciences, 32-46 Aleja Lotnik\'{o}w, 02-668 Warsaw, Poland}

\ead{keesvanderbeek@polytechnique.edu}

\begin{abstract}
High energy electron irradiation is used to controllably introduce atomic-scale point defects into single crystalline  Ba(Fe$_{1-x}$Co$_{x}$)$_{2}$As$_{2}$, Ba(Fe$_{1-x}$Ni$_{x}$)$_{2}$As$_{2}$, and BaFe$_{2}$(As$_{1-x}$P$_{x}$)$_{2}$. The appearance of the collective pinning contribution to the critical  current density in  BaFe$_{2}$(As$_{1-x}$P$_{x}$)$_{2}$, and the magnitude of its enhancement in Ba(Fe$_{1-x}$Co$_{x}$)$_{2}$As$_{2}$, conform with the hypothesis of quasi-particle scattering by Fe vacancies created by the irradiation. Whereas the insignificant modification of the temperature dependence of the superfluid density in Ba(Fe$_{1-x}$Co$_{x}$)$_{2}$As$_{2}$ and Ba(Fe$_{1-x}$Ni$_{x}$)$_{2}$As$_{2}$ points to important native disorder present before the irradiation, the critical temperatures of these materials undergo a suppression equivalent to that observed in the much cleaner BaFe$_{2}$(As$_{1-x}$P$_{x}$)$_{2}$. This  lends credence to the hypothesis of line nodes of the order parameter (at finite $k_{z}$) in the former two materials.
\end{abstract}


\section{Introduction}
The premise of $s_{\pm}$ superconductivity in the multiband iron-based superconductors \cite{Singh2008,Mazin2008}, with a sign-changing order parameter between the electron-like and hole-like Fermi-surface sheets \cite{Mazin2008,Kuroki2008,Kuroki2008ii}, has raised the question of the effect of atomic-scale point-like disorder on superconductivity in these materials \cite{Onari2009,Kontani2010}. In particular,  interband scattering of quasiparticles by point-like disorder was suggested to be pair-breaking, and, therefore, to lead to a strong suppression of the critical temperature, the appearance of sub-gap states \cite{Glatz2010}, and a peculiar $T^{2}$ dependence of the London penetration depth at low temperature \cite{Martin2009,Kogan2009,Gordon2009,Gordon2009ii,Martin2010,Gordon2010,Prozorov2011}. 

More recently, the question of the role of point-like disorder has become only more relevant \cite{Hirschfeld2011}. This is because the anti-ferromagnetic coupling between Cooper pairs on different Fermi surface sheets  \cite{Mazin2008,Kuroki2008,Kuroki2008ii}, and the subtle changes of the Fermi surface with doping \cite{Kuroki2008,Kuroki2008ii}, isovalent chemical substitution \cite{Jiang2009,Wang2009,Thirupathaiah2011,Yoshida2011,Suzuki2011,Putzke2012,Hashimoto2012}, and pressure \cite{Andersen2011,Ahilan2009,Colombier2009,Hassinger2012}, allow for a wide variation of realizations of the order parameter \cite{Kuroki2008,Kuroki2008ii,Hashimoto2012}. In particular, the likelihood of line nodes of the order parameter in 
isovalently doped BaFe$_{2}$(As$_{1-x}$P$_{x}$)$_{2}$ \cite{Hashimoto2010}, LaFePO \cite{Fletcher2009}, and KFe$_{2}$As$_{2}$ \cite{Hashimoto2010ii,Reid2012ii,Maier2011} are indicative of a nodal $s_{\pm}$-- or a $d$-wave superconductivity  \cite{Kuroki2008,Kuroki2008ii} in the first two and latter respective cases. 

Nodal gap structure not imposed by symmetry, but arising from {\em e.g.}, the anisotropy of the Fermi surface and/or interband scattering  \cite{Maier2009}, has  also been evoked \cite{Mishra2009}. In the case of an $s_{\pm}$ state,  the  order parameter sign would not only change between bands but also within a given band  \cite{Mishra2009}. One can also have an $s$--wave gap-structure  in which nodes appear for non-zero $k_{z}$ \cite{Hirschfeld2011,Graser2010, Hirschfeld2010}, a situation that might be relevant for  Ba(Fe$_{1-x}$Co$_{x}$)$_{2}$As$_{2}$ \cite{Reid2010}, Ba(Fe$_{1-x}$Ni$_{x}$)$_{2}$As$_{2}$ \cite{Martin2010}, 
Na$_{1-\delta}$(Fe$_{1-x}$Co$_{x}$)As \cite{Cho2012},  and overdoped Ba$_{1-x}$K$_{x}$Fe$_{2}$As$_{2}$  \cite{Reid2012}.

\subsection{Disorder effect on exotic superconductivity in iron pnictides}
Given that the appearance of gap nodes in an $s$-wave superconductor 
does not imply symmetry breaking, indirect experiments yield, as yet, most insight into the order parameter structure. Such experiments include  the low temperature behaviour of the penetration depth \cite{Martin2009,Kogan2009,Gordon2009,Gordon2009ii,Martin2010,Gordon2010,Prozorov2011,Hashimoto2012,Fletcher2009,Hashimoto2010} and the thermal conductivity \cite{Reid2012ii,Hashimoto2010,Reid2010,Reid2012} as probes of the low-energy quasi-particle density of states (QPDOS),   
but also the sensitivity of the iron-based superconductors  to quasi-particle scattering by homogeneous atomic-scale point-like disorder. This can be introduced by, {\em e.g.}, chemical substitution or energetic particle irradiation. Namely, the occurrence of sign changes of the order parameter implies that even  non-magnetic point-like scatterers may act as pair-breakers \cite{Onari2009,Kontani2010,Glatz2010,Hirschfeld2011}. For  $s_{\pm}$--superconductivity in particular, impurity scattering of quasi-particles between electron--like and hole--like bands (with opposite sign of the order parameter) would lead to a rapid suppression of the critical temperature $T_{c}$  \cite{Onari2009,Kontani2010,Glatz2010}, and open the possibility for non-sign changing $s_{++}$ superconductivity due to orbital fluctuations to manifest itself \cite{Onari2009,Kontani2010,Efremov2012}. The vulnerability of the $d$--wave symmetry to both interband- and intraband impurity scattering would render this even more fragile than $s_{\pm}$.  On the other hand, a less pronounced role of interband  compared to interband scattering would reduce the sensitivity of $s_{\pm}$ superconductivity to disorder \cite{Hirschfeld2011,Efremov2012}. Finally,  the presence of ``accidental'' nodes would result in a rather high initial sensitivity of superconductivity to disorder, with a slower suppression of the critical temperature as function of disorder strength once the interband scattering is sufficiently strong to wash out, or ``lift'' the nodes on all Fermi surface sheets \cite{Mishra2009}.

\begin{table}[b]
\caption{\label{scattering} Contribution of dopant disorder to elastic scattering parameters of various iron pnictide superconductors, such as estimated from the weak collective pinning contribution to the (flux pinning) critical current density, $j_{c}^{coll}$. Here, $k_{F}$ is the Fermi wavevector, $\xi_{0}$ is the Bardeen-Cooper-Schrieffer coherence length, $n_{d}$ is the atomic point defect density, $\sigma_{tr}$ is the transport scattering cross-section, $\delta_{0}$ is the scattering phase angle,  $\Gamma =  n_{d} [ \pi N(0) ]^{-1} \sin^{2} \delta_{0}$ is the scattering rate, where $N(0) = m k_{F} / 2 \pi \hbar^{2}$ is the normal-state DOS.} 
\small
\begin{center}
\lineup
\begin{tabular}{lccccccccc}
\br                              
\em  Material & impurity &$k_{F}$ &  $\xi_{0}$  & $n_{d}$ & $\sigma_{tr}$  & $n_{d}\xi_{0}^{3}$ &  $\sin \delta_{0}$ & $\Gamma$  \rm \cr
& & \AA$^{-1}$ & nm & nm$^{-3}$ &  \AA$^{2}$ & & & meV \cr
\mr
PrFeAsO$_{1-y}$   & O vacancy                    & 0.33 & 2.4 & 1.5 & 6.7     & 21 &  0.3(2) &  10  \cr
NdFeAsO$_{0.9}$F$_{0.1}$  \protect\cite{vdBeek2010}    &     F         & 0.33 & 3.3 & 1.5 & 2.5    & 54 &  0.2 & 4   \cr
Ba$_{0.72}$K$_{0.28}$Fe$_{2}$As$_{2}$  \protect\cite{Wang2010} & K & 0.4  & 2.4 &  2.8  &  1.5   & 38  &  0.1(4) & 3  \cr
Ba$_{0.6}$K$_{04}$Fe$_{2}$As$_{2}$  \protect\cite{Yang2008} &K   & 0.5  & 2.2 &  4  & $2.5 \pm 1.3$   & 43  &  0.2 & 8  \cr
Ba$_{0.45}$K$_{0.55}$Fe$_{2}$As$_{2}$ \protect\cite{vdBeek2010ii}   & K  & 0.5  & 2.2 &  5.5  &    1.5  & 59  &  0.2 &  10   \cr
Ba(Fe$_{0.95}$Co$_{0.05}$)$_{2}$As$_{2}$  \protect\cite{vdBeek2011}& Co & 0.25 & 1.6 &  1  & 2.5 & 8  &  0.17 & 5   \cr
Ba(Fe$_{0.9}$Co$_{0.1}$)$_{2}$As$_{2}$  \protect\cite{Yamamoto2009} & Co & 0.25 & 1.6 &  2  & 2.5    & 8  &  0.17 & 5   \cr
 Ba(Fe$_{0.76}$Ru$_{0.24}$)$_{2}$As$_{2}$  \protect\cite{Ru} & Ru  & 0.25 & 1.6 &  4.8  & 2.5   & 8  &  0.17 & 5   \cr
BaFe$_{2}$(As$_{0.67}$P$_{0.33}$)$_{2}$ \protect\cite{vdBeek2010ii} & P & 0.3 \protect\cite{Shishido2010} & 1.6 & 3.3 & 14 &  --  & -- & -- \cr

\br \normalsize
\end{tabular}
\end{center}
\end{table}

\subsection{Effect of native disorder}
Even if the problem at hand is characterized by a multitude of parameters, it is important to at least establish trends as controlled point--like disorder is added to iron-based superconductors.  Early experiments have suggested a rather limited sensitivity of superconductivity in the iron-based materials to disorder. For example, the residual electronic term in the specific heat suggests a large low--temperature QPDOS \cite{Hardy2010,Mu2010} away from optimal doping, that may be due to pair-breaking. The ubiquitous $T^{2}$--behaviour of the low--temperature $ab$--plane penetration depth $\lambda_{ab}(T)$ suggest strong pair-breaking \cite{Martin2009,Kogan2009,Gordon2009,Gordon2009ii,Martin2010,Gordon2010,Prozorov2011}. Still, $T_{c}$'s of the materials in question are high. 

\begin{figure}[h]
\begin{minipage}{0.75\textwidth}\hspace{-0.8cm} 
\vspace{-4mm}
\includegraphics[width=1.35\textwidth]{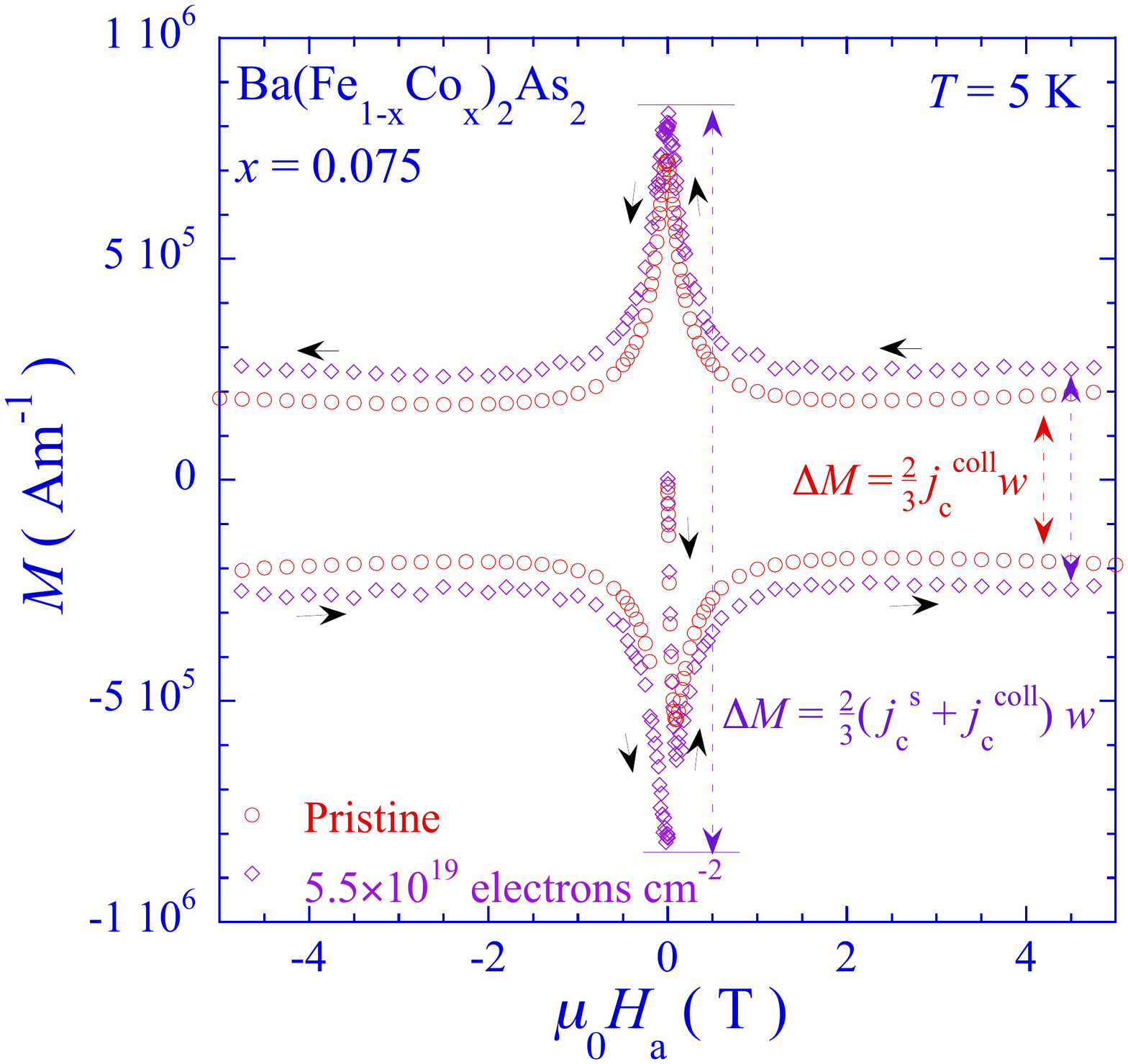}
\end{minipage}~\begin{minipage}[h]{9.5pc}
\caption{ \label{fig:loop} Single crystalline Ba(Fe$_{0.925}$Co$_{0.075}$)$_{2}$As$_{2}$: Hysteresis loop of the irreversible magnetization, at 5~K, before (\color{red}$\circ$\color{black}) and after (\color{RoyalPurple}$\diamond$\color{black})  irradiation with $5.5\times 10^{19}$ cm$^{-2}$ 2.5 MeV electrons.  Closed arrows depict the direction in which the loop is traversed upon cycling the magnetic field. Dotted double arrows depict the width of the magnetization loop in the low--field, strong pinning regime, and in the higher field regime in which only the weak collective pinning contribution $j_{c}^{coll}$ is relevant.}
\end{minipage}
\end{figure}

In our earlier work \cite{vdBeek2010,vdBeek2010ii}, we have drawn attention to the fact that vortex pinning in the mixed state of the iron-based superconductors is indicative of the importance of native disorder. While the critical current density maximum $j_{c}$ at low fields, and the subsequent $j_{c} \sim B^{-1/2}$ drop--off are due to nm-scale heterogeneities of the superconducting properties, leading to a so-called ``strong--pinning'' contribution $j_{c}^{s}$ to the critical current (see Fig.~\ref{fig:loop} and Refs.~\cite{Demirdis2011,vdBeek2011}), the so-called ``weak collective pinning''  \cite{Larkin79,Blatter94} contribution $j_{c}^{coll}$ to the critical current density in the magnetic field range of several tenths of a T to several T  is consistently interpreted in terms of  quasi-particle scattering by the dopant atoms in charge--doped iron-based compounds.  Estimates of scattering rates of the doping impurities, such as these are obtained from flux pinning, yield rather large values, which are at odds with a superconducting ground state that would be sensitive to point-like disorder should this be pair-breaking.  Moreover, it may be noted that the  occurrence of the quasi-particle scattering contribution to flux pinning consistently coincides with that of the $T^{2}$ behaviour of the penetration depth. Inversely,  it is conspicuously absent in clean materials with a $T$-linear dependence of $\lambda_{ab}$, such as  BaFe$_{2}$(As$_{1-x}$P$_{x}$)$_{2}$ \cite{Hashimoto2010,vdBeek2010ii,Demirdis2012}.

The relative insensitivity of superconductivity in the iron-based materials may be due to these materials not having a sign-changing order parameter, but  may also be attributed to limited interband scattering by the dopant impurities. Given the premise of Coulomb scattering (with a small change in crystal momentum) by the charged dopants \cite{vdBeek2010ii}, this is not unreasonable. Estimates from pinning situate the native point defects rather in the Born limit (see table~\ref{scattering}).

\subsection{Artificial disorder}
An obvious manner to test the above ideas is through the artificial introduction of different kinds of point-like defects, either by chemical doping \cite{Cheng2010,Li2011,Li2012}, or by irradiation \cite{Tarantini2010,Nakajima2010}.  An example is the substitution of magnetic ions, which  suppresses $T_{c}$ rather more effectively than non-magnetic substitutions \cite{Cheng2010,Li2012}. A major drawback of chemical substitution in the iron-based compounds is, however, that this simultaneously leads to structural changes and/or doping. Energetic particle irradiation therefore seems preferable. Recent work by Tarantini {\em et al.} reported on $T_{c}$--suppression by  $\alpha$--particle irradiation of NdFeAs(O,F), but was prone to criticism, notably in that it induces magnetic impurities and defect clustering. The 3 MeV proton irradiation experiments by Nakajima {\em et al.} on Ba(Fe$_{1-x}$Co$_{x}$)$_{2}$As$_{2}$ reported monotonous $T_{c}$--depression as the residual resistivity increases; the authors provided estimates for the critical pair-breaking parameter as  $\Gamma / 2 \pi T_{c} \sim 4 - 7 $ and $\Gamma / 2 \pi T_{c} \sim 1.5 - 2.5 $, as one goes from underdoped to overdoped \cite{Nakajima2010}. However,  apart from point defects, proton and neutron irradiation are known to induce, {\it in situ}, point defect cascades and clusters, which may play a different role than that of simple scatterers. 

Here,  we compare  the effect of 2.5 MeV electron irradiation on differently substituted BaFe$_{2}$As$_{2}$. Such high--energy electron irradiation is known to produce homogeneously distributed vacancy--interstitial (Frenkel) pairs on all sublattices of a crystal, but, preferentially on that constituted by the lightest atomic species in the material -- in the materials studied here, Fe. In the high-$T_{c}$ cuprates, electron-irradiation defects are known to be strong unitary scatterers, comparable to Zn substitution. There, they are responsible for the suppression of the critical temperature \cite{Albenque2000,Albenque2003,Spathis2008}, and the appearance of a $T^{2}$--temperature dependence of the penetration depth \cite{Spathis2008}. In semi-metals with extremely low carrier densities, such as Bi, the irradiation-induced defects are charged and lead to effective doping of the material \cite{Beuneu87}. The mobility of the created point defects at high temperature is a caveat, since it leads to partial annealing and clustering, albeit it to a lesser extent than that produced by proton or neutron irradiation.

It turns out that electron irradiation leads to similar monotonic suppression of $T_{c}$ in Ba(Fe$_{1-x}$Co$_{x}$)$_{2}$As$_{2}$,  Ba(Fe$_{1-x}$Ni$_{x}$)$_{2}$As$_{2}$, and  BaFe$_{2}$(As$_{1-x}$P$_{x}$)$_{2}$. In Ba(Fe$_{1-x}$Co$_{x}$)$_{2}$As$_{2}$, the suppression is weakest for the optimally doped material and strongest for underdoped crystals.  In the range of intermediate electron fluences studied here,  the resistance monotonically increases as function of fluence. In Ba(Fe$_{1-x}$Co$_{x}$)$_{2}$As$_{2}$ and  Ba(Fe$_{1-x}$Ni$_{x}$)$_{2}$As$_{2}$, there is little to no change of the temperature dependence of the superfluid density. We have also measured the critical current density before and after the irradiation. In Ba(Fe$_{1-x}$Co$_{x}$)$_{2}$As$_{2}$ one has a clear increase of the weak collective pinning contribution to $j_{c}$, while in BaFe$_{2}$(As$_{1-x}$P$_{x}$)$_{2}$ the initially absent weak pinning contribution emerges after irradiation. This finding establishes that the field--independent contribution $j_{c}^{coll}$ to the critical current density of iron-based superconductors is, indeed, due to atomic sized point pins.  In both materials, this finding allows for an estimate of the density of produced defects per unit irradiation fluence. In  Ba(Fe$_{1-x}$Ni$_{x}$)$_{2}$As$_{2}$, $j_{c}$ is  strongly suppressed by the irradiation. 

\begin{figure}[h] \hspace{0cm}
\includegraphics[width=1.0\textwidth]{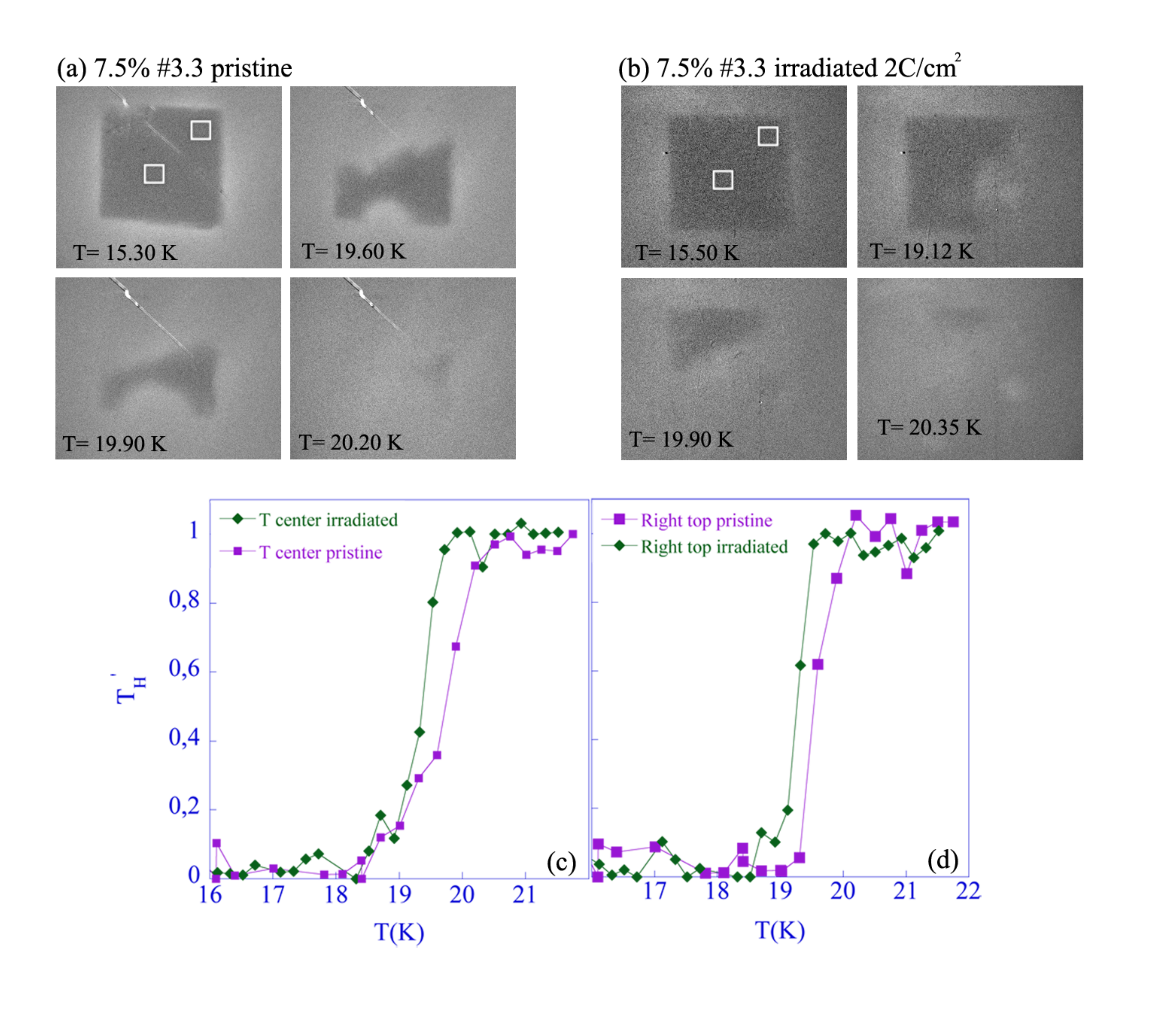}
\vspace{-15mm}
\caption{\label{fig:transmittivity+dmo} Transition from the superconducting to the normal state of a Ba(Fe$_{0.925}$Co$_{0.075}$)$_{2}$As$_{2}$ single crystal before (a,c) and after irradiation with 2.1 Ccm$^{-2}$ 2.5 MeV electrons (b,d), as imaged by the differential magneto-optical  (DMO) technique \protect\cite{Demirdis2011}. The images (a,b) show the progressive admission of an ac magnetic field of magnitude 1~Oe, applied perpendicularly to the sample surface, as the temperature is raised. In these flux density maps, areas of high luminous intensity $I$ correspond to the value of the applied field, while dark areas correspond to zero field, ({\it i.e.}  full screening). The dark rectangles in the upper left hand panels of (a) and (b) correspond to the sample outline, {\em i.e.}, full Meissner expulsion of the magnetic field at the lowest temperature. The white squares indicate the positions at which the ``transmittivity''  (or ``local ac susceptibility'')  defined as $[I(T)-I(T \ll T_{c}) ] /[ I(T\gg T_{c})-I(T \ll T_{c}) ]$ depicted in the lower panels (c,d) was determined. }
\end{figure}

\section{Experimental details}
\subsection{Single crystalline samples}
Single-crystals of Ba(Fe$_{1-x}$Co$_{x}$)$_{2}$As$_{2}$  \cite{Rullier-Albenque2009}  and Ba(Fe$_{1-x}$Ni$_{x}$)$_{2}$As$_{2}$  \cite{Olariu2011} were grown using the self-flux method. The high purity starting reagents Ba, FeAs, and CoAs were mixed in the molar ratio $1:(4-x):x$, loaded in alumina crucibles and then sealed in evacuated quartz tubes. For each doping level, chemical analysis by an electron probe was performed on several crystals yielding the dopant content within 0.5\% absolute accuracy. For this work we studied six Co--doping levels and two Ni--doping levels. 
The optimally doped BaFe$_{2}$(As$_{1-x}$P$_{x}$)$_{2}$ single crystals, with $x = 0.33$ and $x = 0.36$, were also grown by the self flux method \cite{Kasahara2010}, and  characterized using x-ray diffraction and energy dispersive x-ray spectroscopy. No impurity phases were found within the experimental limits of  $\sim \%$1. 

\subsection{Electron irradiation}
The 2.5 MeV electron irradiation was performed at the SIRIUS Pelletron facility of the Laboratoire des Solides Irradi\'{e}s (LSI) at the Ecole Polytechnique in Palaiseau, France \cite{sirius}. Beam currents varied between 10 and 20 $\mu$A, depending on the irradiation run; the beam was swept over a $6\times6$ mm$^{2}$ area. Samples of different composition irradiated to the same dose were mounted together and irradiated simultaneously. In order to prevent {\it in-situ} defect migration, recombination, and clustering, the irradiations were carried out in a liquid $H_{2}$ bath ($T = 20$~K). In--situ resistance measurements were carried out on the Ba(Fe$_{1-x}$Co$_{x}$)$_{2}$As$_{2}$ material in order to assess the increase of the residual resistivity as a function of defect density at low temperature, as well as the effect of room-temperature annealing.

\begin{figure}[t] \hspace{0cm}
\includegraphics[width=1.0\textwidth]{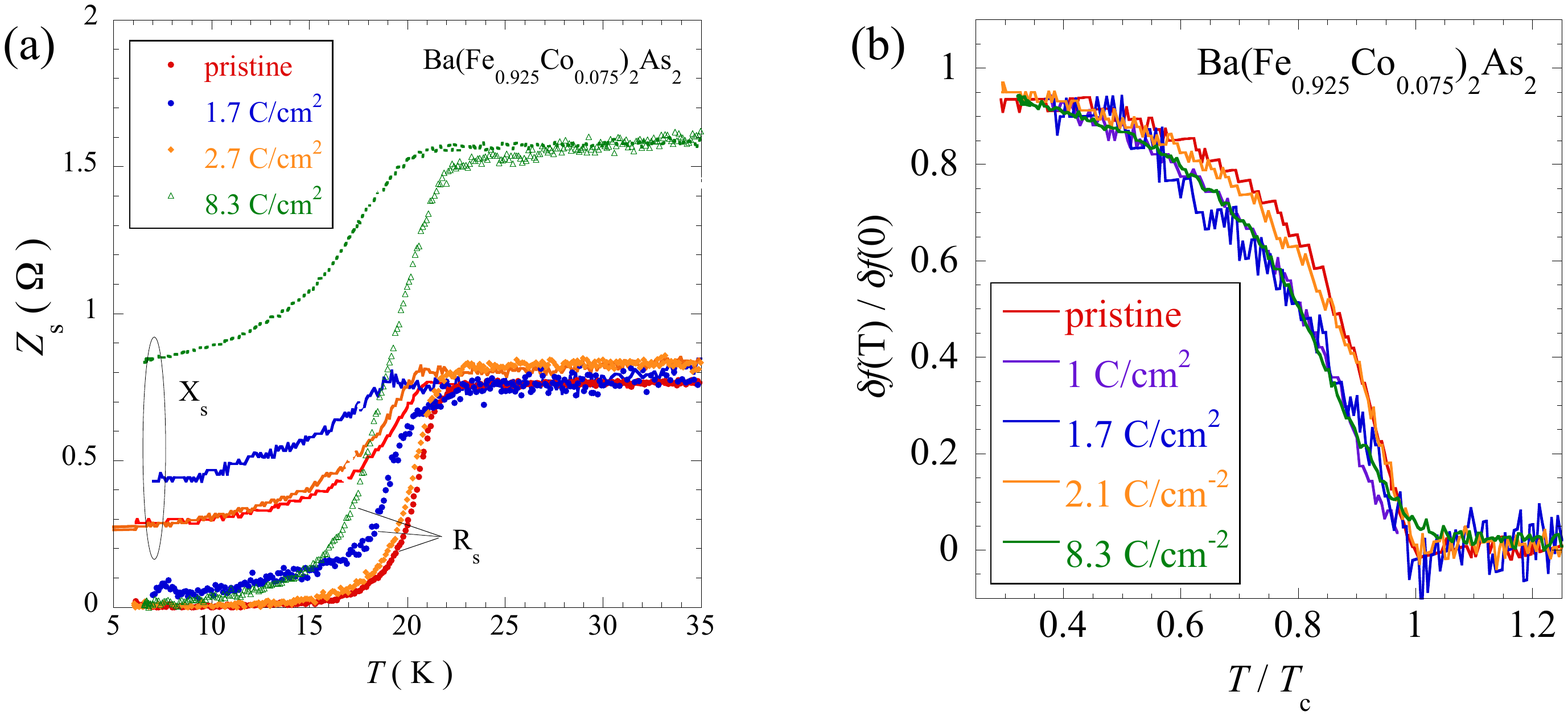}
\vspace{-4cm}
\caption{\label{fig:surface-resistance} 
Surface impedance  $Z_{s}$ of Ba(Fe$_{0.925}$Co$_{0.075}$)$_{2}$As$_{2}$ crystals before and after various low--temperature irradiation runs with 2.5 MeV electrons, and subsequent annealing at 300~K. Data points and thin lines show the surface resistance $R_{s}$ and reactance $X_{s}$ respectively, for various electron doses. (b) Shift $\delta f(T)$ of the resonant frequency of the superconducting Nb cavity (operating at 26.0 GHz) as function of temperature, normalized with respect to the low--temperature extrapolated $\delta f(0)$, for Ba(Fe$_{0.925}$Co$_{0.075}$)$_{2}$As$_{2}$ crystals irradiated with various fluences of 2.5 MeV electrons. }
\end{figure}

\subsection{Measurements of $T_{c}$ and the surface impedance}
$T_{c}$--values before and after irradiation were determined using differential magneto-optical (DMO) imaging (see Fig.~\ref{fig:transmittivity+dmo} and Ref.~\cite{Demirdis2011}). For the lower irradiation doses, in particular, the shift in $T_{c}$ after irradiation is smaller than the transition width. DMO then provides an objective means to quantify the effect of the electron irradiation on different portions of a crystal. The change in $T_{c}$ was further quantified using {\it ex-situ} measurements of the temperature dependent  microwave surface impedance $Z_{s} = R_{s} + i X_{s}$. These were performed at the LSI using a superconducting Nb cavity, operating at 26.0 GHz, and cooled to 5~K using a 0.5~W cryocooler cold head. From the temperature dependent resonance frequency $f$ and unloaded quality factor $Q_{0}$, the surface resistance $R_{s}$ and reactance $X_{s}$ were determined as 
\begin{equation}
R_{s} = \mu_{0} f G  \left[ \frac{1}{Q_{0}(T)} - \frac{1}{\tilde{Q}_{0}} \right] 
\end{equation}
and 
\begin{equation}
X_{s}(T) =   \mu_{0} G \left[ f(T)  - f(0) \right]  + X_{0}
\end{equation}
respectively. Here $f(0)$ and $\tilde{Q}_{0}$ are the resonance frequency and the unloaded quality factor in the absence of the sample, and the geometrical factor $G = V / 4\pi w \sqrt{wd}$, with $V$ the inner volume of the cavity, $d$ the thickness of the platelet-like single crystal sample, and $w$ its smaller width. The additive constant $X_{0}$ was adjusted so that $R_{s} = X_{s}$ in the normal state. 

\subsection{Determination of the critical current density}
Measurements of the magnetization $M$ of the Ba(Fe$_{1-x}$Co$_{x}$)$_{2}$As$_{2}$ and Ba(Fe$_{1-x}$Ni$_{x}$)$_{2}$As$_{2}$ single crystals as function of the applied magnetic field $H_{a}$ were performed using a commercial SQUID magnetometer. From the width $\Delta  M(H)$ of the magnetic hysteresis loops, the field-and temperature dependence of the sustainable current density $j = \frac{3}{2} \Delta M / w$ was determined using the Bean model. The numeric prefactor  $\frac{3}{2}$ is estimated from  calculations of E.H. Brandt for rectangular bars of similar aspect ratio as the measured crystals \cite{Brandt96}. Data on  BaFe$_{2}$(As$_{1-x}$P$_{x}$)$_{2}$ were acquired by M. Konczykowski using the local Hall probe magnetometry technique \cite{Konczykowski2011,Konczykowski2012}.  Briefly, the single crystalline superconducting crystal is centered on top of an array of 11 Hall sensors, of active area $3 \times 3$ $\mu$m$^{2}$, and spaced by 20~$\mu$m, in such a manner that  the short er crystal edge is parallel to the array, and the array spans the longer crystal edge. The Hall sensors are used to measure the spatial gradient $dB/dx$ of the local induction component perpendicular to the crystal surface. The gradient $dB/dx$ ( in G/$\mu$m ) is directly proportional to the sustainable screening current $j$, with $dB/dx \sim \frac{1}{6} j$ for a crystal aspect ratio of $\sim 0.1$ \cite{Brandt96}. Measurements of the flux density gradient were performed both as function of $H_{a}$ after zero field--cooling, and as function of temperature $T$, after field cooling, and the subsequent reduction of $H_{a}$ by twice the full penetration field. 

In what follows, we shall take the screening current at the lowest measurement temperature ( 5~K ) to be representative of the critical current density. At higher temperature, the measured current density can be significantly lower than the pinning critical current density because of flux creep \cite{Demirdis2012,Konczykowski2012}.

\section{Analysis of the critical current density} 
The critical current density is further analyzed along the lines of Refs.~\cite{vdBeek2010,vdBeek2010ii}. The hysteresis loops of all charge--doped iron-based superconductors, exemplified by Fig.~\ref{fig:loop}, indicate the presence of two main pinning mechanisms in these materials. The central peak of the magnetization, at low applied fields, is due to the strong--pinning contribution $j_{c}^{s}$ to the critical current, while the constant contribution at higher fields comes from ``weak collective pinning'' by the dopant atoms. In a critical current density - versus magnetic field plot (see Fig.~\ref{fig:jc-dose}) the strong pinning contribution is responsible for the low--field plateau and the subsequent $j_{c} \propto B^{-1/2}$ decrease, while the collective pinning contribution yields the high--field constant $j_{c}$.  It can be written in terms of superconducting parameters, defect density  $n_{d}$, and the elementary pinning force $f_{p}$ of a single defect, as 
\begin{equation}
j_{c}^{coll} = j_{0} \left( \frac{n_{d}\langle f_{p}^{2} \rangle \xi_{ab}^{3}}{\varepsilon_{\lambda}\varepsilon_{0}} \right)^{2/3} .
\label{eq:jc-coll}
\end{equation}
Here, $j_{0}$ is the depairing current density, $\xi_{ab}$ and $\lambda_{ab}$  are the $ab$--plane coherence length and penetration depth respectively, $\varepsilon_{0} \equiv \Phi_{0}^{2} / 4 \pi \mu_{0} \lambda_{ab}^{2} $ is the vortex line energy, $\varepsilon_{\lambda}$ is the penetration depth anisotropy \cite{vdBeek2012}, and $\Phi_{0} = h/2e$ is the flux quantum. The averaging $\langle \ldots \rangle$ is performed over the vortex core. 

The magnitude of $j_{c}^{coll}$ is compatible with scattering of the quasi-particles in the vortex cores as being at the origin of the weak collective pinning contribution in pristine iron-based superconductors, provided that the dopant atoms are the scattering defects. This gives rise to  \cite{Thuneberg82,Thuneberg84}
\begin{equation}
f_{p} = 0.3 g(\rho_{D}) \varepsilon_{0} \left( \frac{\sigma_{tr}}{\pi \xi_{ab}^{2}}\right) \left( \frac{\xi_{0}}{\xi_{ab}}\right)
\label{eq:Thuneberg}
\end{equation}
which depends not only on the Gor'kov impurity parameter $\rho_{D} = \hbar v_{F} / 2\pi T_{c} l = \xi_{0} / l$, but also on the transport cross-section $\sigma_{tr}  = (2\pi/k_{F}^{2})\sin^{2} \delta_{0}$ ($\delta_{0}$ is the scattering phase angle, and $k_{F}$ the Fermi wavevector). The analysis of the intermediate field critical current density of  several common iron based superconductors allows one to estimate the scattering parameters compiled in Table~\ref{scattering}. The weak collective pinning contribution is absent in isovalently doped BaFe$_{2}$(As$_{1-x}$P$_{x}$)$_{2}$ for all doping levels $x$ [see Ref.~\cite{Demirdis2012} and Fig.~\ref{BaFeAsP-loops}(a)]; surprisingly, it is present in  Ba(Fe$_{1-x}$Ru$_{x}$)$_{2}$As$_{2}$ \cite{Ru}  and LiFeAs \cite{Konczykowski2011}.

\begin{figure}[t] \hspace{0cm}
\begin{minipage}[h]{0.7\textwidth}\hspace{-5mm}
\includegraphics[width=1.1\textwidth]{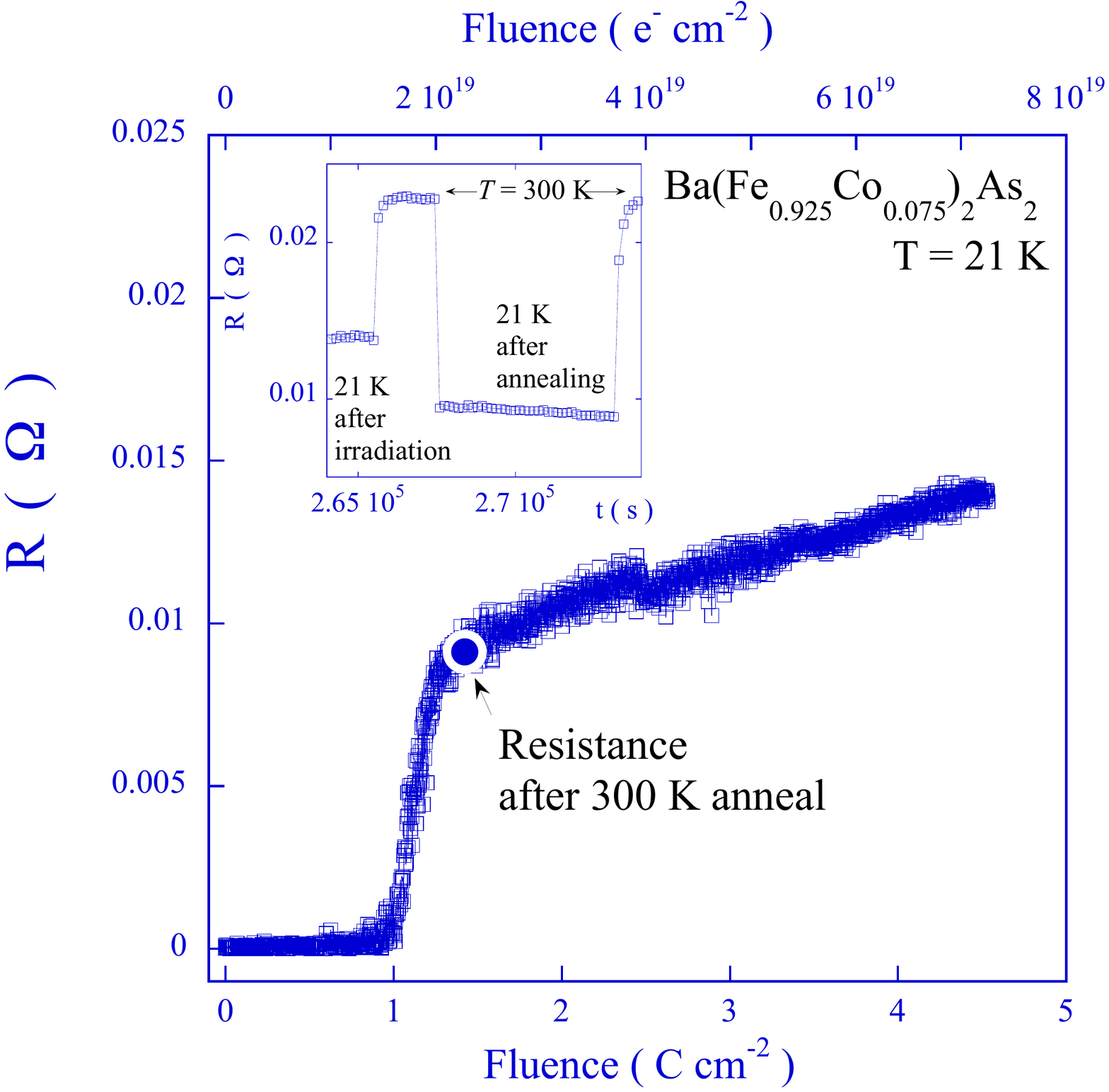}
\end{minipage}~\hspace{-17mm}\begin{minipage}[h]{0.4\textwidth}
\caption{\label{fig:Annealing} Ba(Fe$_{0.925}$Co$_{0.075}$)$_{2}$As$_{2}$: Resistance as function of dose of 2.5 MeV electrons, measured at 21 K.  The sample transits to the normal state after a dose of 1.2 Ccm$^{-2}$, after which the resistance increases at a rate of $\delta R / R = 0.14 $[Ccm$^{-2}$]$^{-1}$. The Inset shows the effect of annealing. The resistance of the crystal is represented as function of time. The protocol comprises initial measurements at 21~K following irradiation with 4.6 Ccm$^{-2}$ 2.5 MeV electrons. The crystal is then warmed to 300~K, cooled to 21~K, and heated once again to 300~K. After the first anneal at 300~K, the resistance drops back to the value reached after low-temperature irradiation with only 1.4 Ccm$^{-2}$ (white circle on the curve in the main panel).  }
\end{minipage}
\label{fig:annealing}
\end{figure}

\section{Results and Discussion}
\subsection{Ba(Fe$_{1-x}$Co$_{x}$)$_{2}$As$_{2}$}
\label{section:Co}
Fig.~\ref{fig:Annealing} shows the resistance of an optimally doped Ba(Fe$_{0.925}$Co$_{0.075}$)$_{2}$As$_{2}$ single crystal, measured   {\em in-situ} during exposure to the 10~$\mu$A, 2.5 MeV beam, at $T = 21$~K.  The sample transits to the normal state at a fluence of 1.2 Ccm$^{-2}$, after which the resistance increases at a rate of $\delta R / R = 0.14 $[Ccm$^{-2}$]$^{-1}$. After a fluence of 4.6 Ccm$^{-2}$ was reached, the electron beam was stopped,  the sample heated to 300~K, and the resistance measured. The crystal was subsequently cooled down once again to 21~K, and the resistance measured again (Inset to Fig.~\ref{fig:Annealing}). It appears that after the 300~K anneal, the resistance has dropped to the value first measured after irradiation with a fluence of 1.4 Ccm$^{-2}$. Therefore, annealing at 300~K leads to the annihilation and clustering of the point defects produced by the irradiation, with a resulting drop of the irradiation--induced resistance change of 65~\%. Since all data presented below concern  {\em ex-situ} experiments performed after heating the crystals to 300~K, the effect of  annealing should be taken into account.

\begin{figure}[t] \hspace{0cm}
\includegraphics[width=1.1\textwidth]{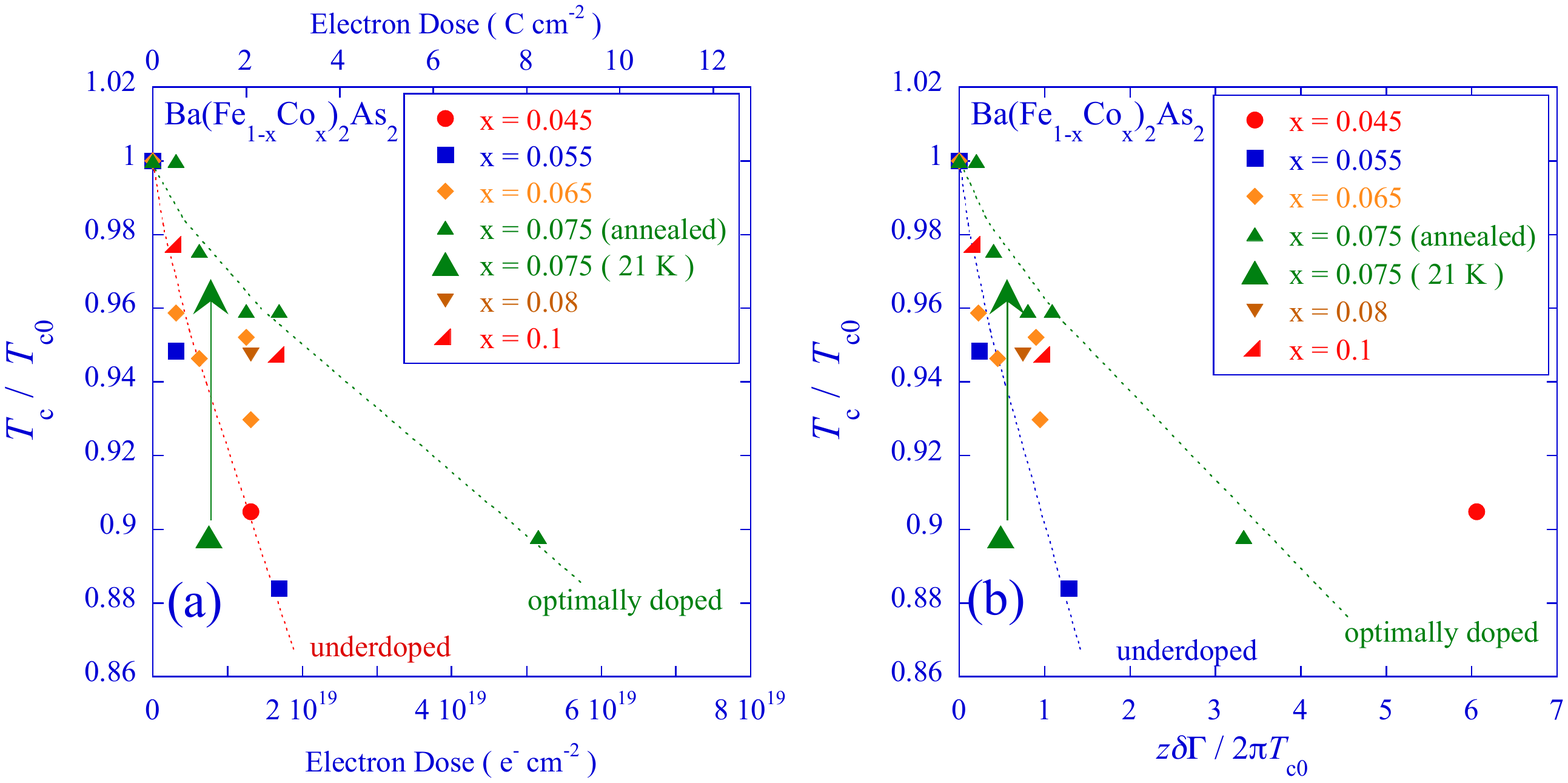}
\vspace{-49mm}
\caption{\label{fig:BaFeCoAs-Tc-vs-dose} Ba(Fe$_{1-x}$Co$_{x}$)$_{2}$As$_{2}$: evolution of  $T_{c}$ (normalized by the initial critical temperature $T_{c0}$) as function of (a) the fluence of 2.5 MeV electrons (b) the estimated normal state scattering rate $\Gamma$ (normalized by $T_{c0}$), for different doping levels $x$. The small data points show the $T_{c}$--values after annealing at 300~K. The large green triangle shows the drop of $T_{c}$ following low-temperature irradiation of optimally doped Ba(Fe$_{0.925}$Co$_{0.075}$)$_{2}$As$_{2}$, the arrow shows the effect of annealing at 300~K.}
\label{fig:Tc-vs-dose} 
\end{figure}

The evolution of  the reduced critical temperature $T_{c}/T_{c0}$ with electron fluence (after 300~K annealing) is shown for various Co-substitution levels $x$ in Fig.~\ref{fig:BaFeCoAs-Tc-vs-dose}(a). Here, $T_{c0}$ is the critical temperature of the crystal before irradiation. It is to be stressed that, given the modest $T_{c}$ changes involved, and the typical dispersion of $T_{c}$--values between similarly grown samples, even from the same batch, it is essential to compare the values of $T_{c}$ and $T_{c0}$ on the same crystal before and after irradiation. For the same reason, a plot of the absolute $T_{c}$ versus dose would be meaningless in the regime of small doses. Fig.~\ref{fig:BaFeCoAs-Tc-vs-dose} also shows the effect of 300~K annealing, for one sample. It is seen that the change in $T_{c}$ drops by a factor three after annealing. 

Fig.~\ref{fig:BaFeCoAs-Tc-vs-dose}(b) shows the same data, as function of the product of the induced change in the scattering rate $\delta \Gamma/2\pi T_{c}$ and the electronic effective mass enhancement  $z = m^{*}/m_{e}$. Here $m_{e}$ is the free electron mass,  and $m^{*}$ the effective mass. $z \delta \Gamma = e \delta \rho / R_{H} m_{e} $ (with $e$ the electronic charge)  is estimated from the change of the resistance due to the irradiation (after annealing), together with the published data for the resistivity $\rho$ and the Hall coefficient $R_{H}$ published in Ref.~\cite{Rullier-Albenque2009}. In spite of the contributions of multiple bands to the conductivity of Ba(Fe$_{1-x}$Co$_{x}$)$_{2}$As$_{2}$, this approach can be justified given that the conductivity contribution from the electron sheets dominates the transport properties \cite{Rullier-Albenque2009}.  Typically,  $\delta R/R \sim 0.05$ [Ccm$^{-2}$]$^{-1}$ for the optimally doped sample (cf Fig.~\ref{fig:annealing}). Note, however, that the resistivity change with electron dose depends on $x$, and that,  in order to produce the ``simulated'' $\delta \Gamma$ values of Fig.~\ref{fig:Tc-vs-dose}b, we use data measured in-situ on crystals with different $x$. Estimates of the scattering rate from critical current measurements (see below) yield qualitatively similar results as those shown in the Figure.. 

The surface impedance of optimally doped Ba(Fe$_{1-x}$Co$_{x}$)$_{2}$As$_{2}$ before and after electron irradiation is depicted in Fig.~\ref{fig:surface-resistance}. A monotonous increase of the surface resistance is observed after irradiation; even though in this series of experiments, the increase was not linear as function of fluence. In particular, the crystal irradiated with 2.1 Ccm$^{-2}$ electrons did not conform to the general trend. The normalized frequency shift of the Nb cavity, which is proportional to the superfluid density $n_{s} \propto \lambda_{ab}^{-2}$ of the superconducting crystals, showed a slight change to a less marked temperature dependence, indicative of disorder having been added to the crystal. The trend is similar to that observed by Hashimoto {\em et al.} in Ba$_{1-x}$K$_{x}$Fe$_{2}$As$_{2}$ crystals with varying degrees of disorder \cite{Hashimoto2009ii}, but is much less marked here.

\begin{figure}[t] \hspace{-1cm}
\includegraphics[width=1.14\textwidth]{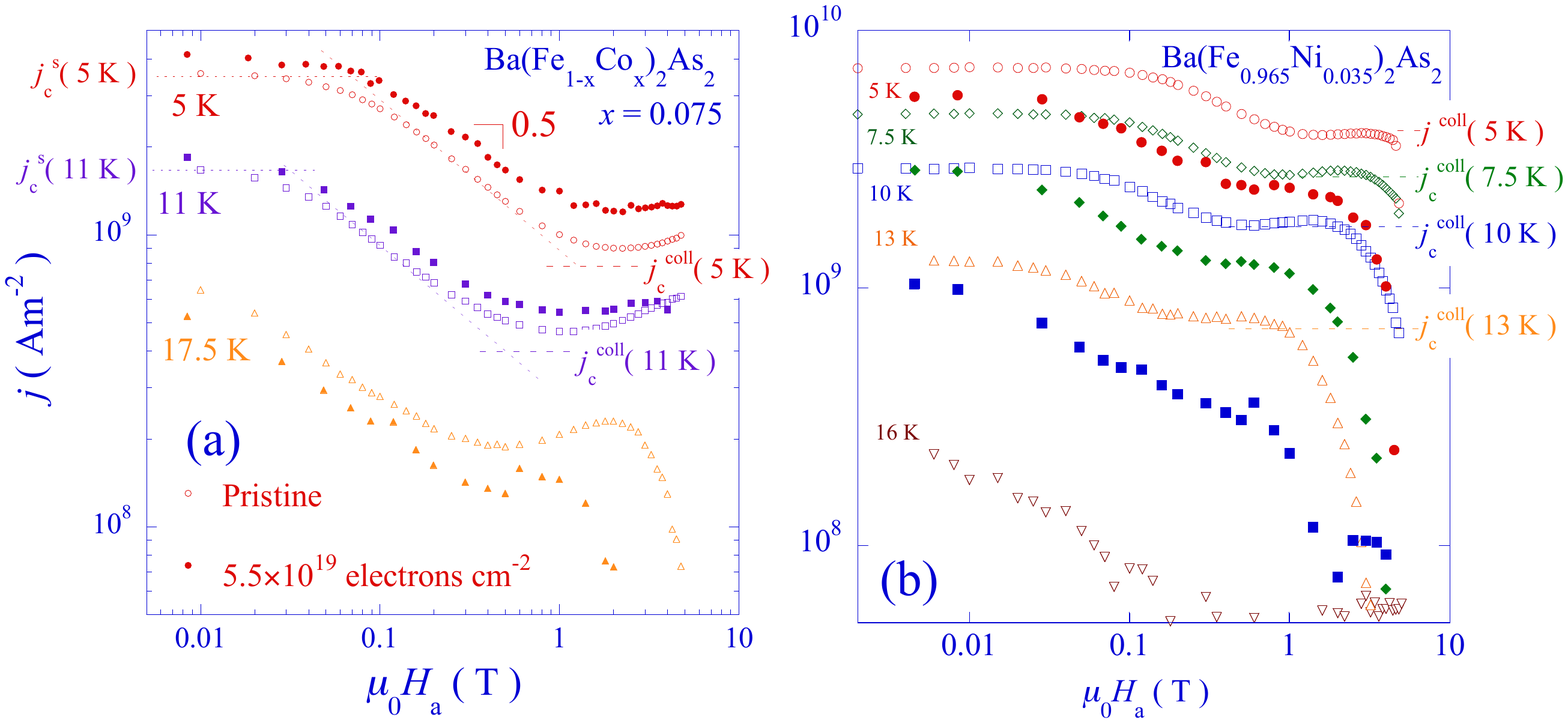}   
\vspace{-55mm}
\caption{\label{fig:jc-dose}  Sustainable current density $j$ of (a) optimally doped Ba(Fe$_{1-x}$Co$_{x}$)$_{2}$As$_{2}$ ($ x = 0.075)$ and (b) optimally doped  Ba(Fe$_{1-x}$Ni$_{x}$)$_{2}$As$_{2}$ ( $x = 0.035$) before (open symbols) and after irradiation at 21 K with $5.5\times 10^{19}$ cm$^{-2}$ 2.5 MeV electrons (closed symbols). The measurements were performed at 5~K ( \color{red} $\circ$ \color{black}, \color{red}$\bullet$ \color{black}), 7.5~K  ( \color{OliveGreen} $\diamond$ \color{black}, \color{OliveGreen} $\diamond$ \color{black}), 10~K ( \color{blue} \protect\raisebox{3pt}{\framebox{\hspace{0.2mm}}} \color{black}, \color{blue} \protect\rule{2.5mm}{2.5mm} \color{black}),  11 K ( \color{RoyalPurple} \protect\raisebox{3pt}{\framebox{\hspace{0.2mm}}} \color{black}, \color{RoyalPurple} \protect \rule{2.5mm}{2.5mm} \color{black}), 13~K (\color{BrickRed} $\triangle$ \color{black}), 16~K (\color{Sepia}$\bigtriangledown$\color{black}), and 17.5~K ( \color{BurntOrange} $\triangle$ \color{black}, \color{BurntOrange} $\triangle$ \color{black}).  }
\end{figure}
 
The low--temperature critical current density of single crystalline  Ba(Fe$_{0.925}$Co$_{0.075}$)$_{2}$As$_{2}$ increases by a constant, field-independent amount after electron irradiation (see Fig.~\ref{fig:jc-dose}a). This indicates that it is the weak collective pinning contribution $j_{c}^{coll}$ that is enhanced, presumably by atomic-sized point defects introduced by the irradiation. If one assumes that the relevant defects are Fe vacancy-interstitial pairs, {\em i.e.} that $\sigma_{tr} \sim \pi D_{Fe}^{2} \sim 2.6$~\AA$^{2}$, with $ D_{Fe}= 0.92$~\AA ~the ionic radius of Fe$^{2+}$, the increase of the critical current density is consistent, through Eqs.~(\ref{eq:jc-coll}) and (\ref{eq:Thuneberg}), with 0.006 dpa Fe / Ccm$^{-2}$.  
The increase of the low--temperature critical current density and the concomitant decrease of $T_{c}$ after irradiation implies a steeper decrease of $j_{c}$ with temperature, and a crossing of $j_{c}$--values of the pristine and the irradiated crystal at an intermediate temperature.

\subsection{Ba(Fe$_{1-x}$Ni$_{x}$)$_{2}$As$_{2}$}
Surface impedance measurements and DMO imaging on the irradiated Ba(Fe$_{1-x}$Ni$_{x}$)$_{2}$As$_{2}$ crystals show a more rapid depression of  $T_{c}$ with dose than in Ba(Fe$_{1-x}$Co$_{x}$)$_{2}$As$_{2}$. If anything, $T_{c}$ versus dose is comparable to that found in  underdoped Ba(Fe$_{1-x}$Co$_{x}$)$_{2}$As$_{2}$ (see Fig.~\ref{Tc-vs-dose}). As in Ba(Fe$_{1-x}$Co$_{x}$)$_{2}$As$_{2}$, the temperature dependence of the superfluid density of Ba(Fe$_{1-x}$Ni$_{x}$)$_{2}$As$_{2}$ changes little or not at all, even at the largest irradiation dose of $5.5 \times 10^{19}$ electrons cm$^{-2}$.

The critical current density of the Ba(Fe$_{1-x}$Ni$_{x}$)$_{2}$As$_{2}$ crystals is strongly suppressed after the irradiation (see Fig.~\ref{fig:jc-dose}b). This is partially because the  reduced measurement temperatures are higher in Ba(Fe$_{1-x}$Ni$_{x}$)$_{2}$As$_{2}$ than in its co-substituted counterpart, but mainly because the suppression of the prefactor $j_{0}  \propto n_{s} e (\Delta / h k_{F} ) $ in Eq.~(\ref{eq:jc-coll}) outweighs the increase in $n_{d}$ ($\Delta$ is the superconducting gap amplitude).

\begin{figure}[t]
\includegraphics[width=\textwidth]{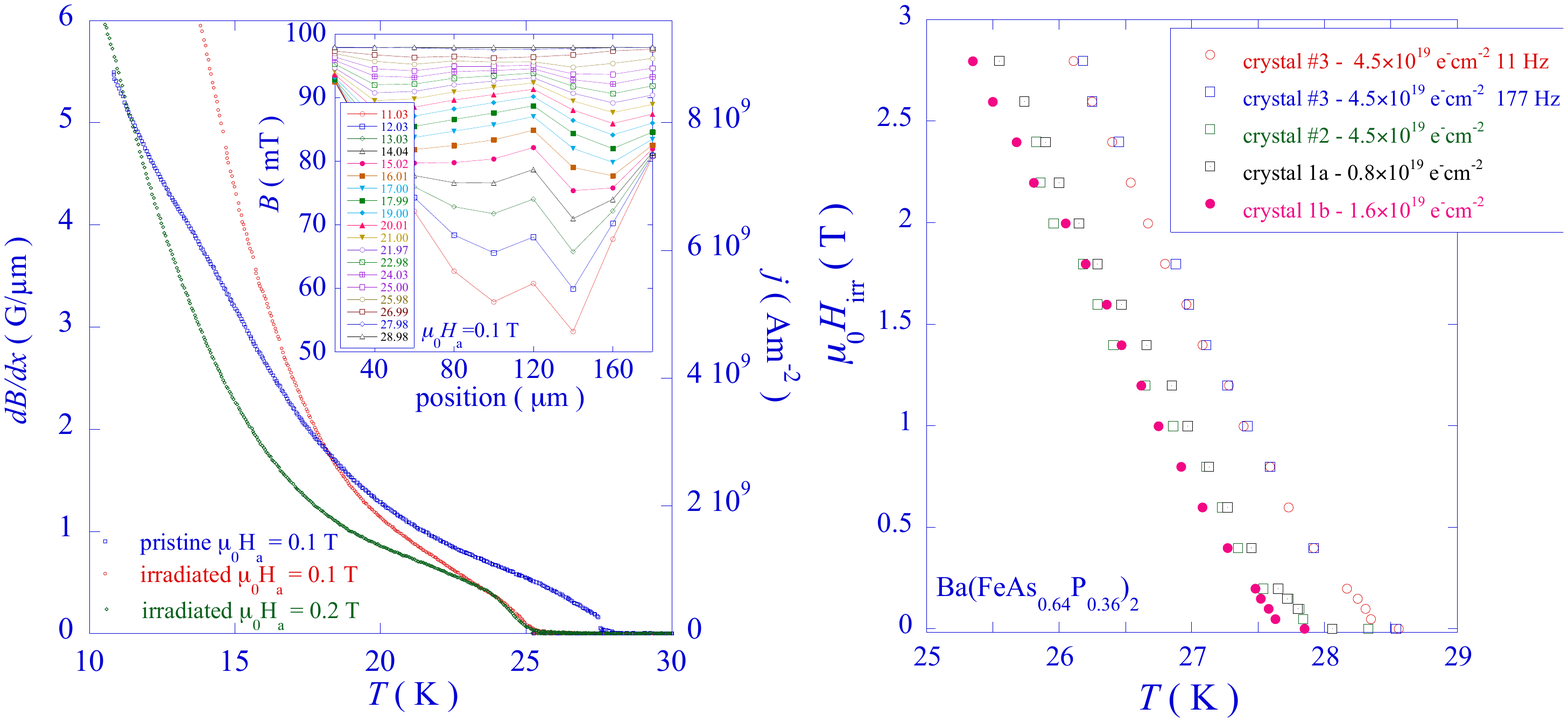}\vspace{-35mm}
 \begin{minipage}[t]{18pc}   \caption{ \label{fig:j-vs-T} Temperature dependence of the flux density gradient $dB/dx$ in a BaFe$_{2}$(As$_{0.64}$P$_{0.36}$)$_{2}$ single crystal before and after irradiation with $0.8\times 10^{19}$ electrons cm$^{-2}$. Inset: Flux density profile across the crystal after zero-field cooling, application of the applied field $\mu_{0}H_{a} = 100$~mT, and subsequent warming. }  \end{minipage}
 \hspace{2pc}%
 \begin{minipage}[t]{18pc}  \caption{\label{fig:IRL} Irreversibility field $BH_{irr}(T)$ as function of temperature, for BaFe$_{2}$(As$_{0.64}$P$_{0.36}$)$_{2}$ single crystals irradiated with $0.8\times10^{19}$ (\protect\raisebox{3pt}{\framebox{\hspace{0.2mm}}}), $1.6\times10^{19}$ (\color{magenta} $\bullet$\color{black}), and $4.5\times10^{19}$ electrons cm$^{-2}$ (\color{red}$\circ$\color{black} , \color{blue}\protect\raisebox{3pt}{\framebox{\hspace{0.2mm}}}\color{black} , \color{OliveGreen}\protect\raisebox{3pt}{\framebox{\hspace{0.1mm}}}\color{black}). $B_{irr}(T)$ is determined as the locus of temperatures $T$ above which the nonlinearity of the crystal current-voltage characteristic vanishes at given induction $H_{a}$. $T_{c}$ is determined from the extrapolation of $H_{irr}(T)$ to zero field. }\end{minipage}
\end{figure}

\subsection{BaFe$_{2}$(As$_{1-x}$P$_{x}$)$_{2}$}
The Inset to Fig.~\ref{fig:j-vs-T} shows an example of flux density gradients measured on a  BaFe$_{2}$(As$_{0.64}$P$_{0.36}$)$_{2}$ single crystal irradiated with $0.9\times10^{19}$ electrons cm$^{-2}$, after zero field cooling and the application of an external field of 100~mT. The temperature dependence of the flux density gradient obtained after field-cooling in 400 mT, reduction of the applied field to the indicated target value, and subsequent warming is shown in the main panel of the Figure. The flux density gradient is directly proportional to the sustainable current density $j$. As in Ba(Fe$_{1-x}$Co$_{x}$)$_{2}$As$_{2}$, the screening current in the irradiated crystal exceeds that of the pristine crystal at low temperature, but drops below it at higher temperature, due to the decrease of $T_{c}$. 

The critical temperatures after electron-irradiation were determined as the extrapolation to zero of the ``irreversibility field'' $H_{irr}(T)$ above which the nonlinearity of the current-voltage characteristic -- and therefore the critical current density -- vanishes. Details on the determination of $H_{irr}$, can be found in, {\em e.g.}, Ref.~\cite{Konczykowski2012}. In all cases, the $H_{irr}(T)$ values depend  very little on the frequency of the ac field used to investigate the screening by the superconducting sample; the corresponding $T_{irr}(H)$ lies very close to the temperatures at which the dc screening current vanishes in Fig.~\ref{fig:j-vs-T}. The  $T_{c}$--values resulting from the extrapolation of $H_{irr}(T)$ to zero are gathered in  Fig.~\ref{Tc-vs-dose}. The $T_{c}$--depression as function of electron fluence is comparable to that in Ba(Fe$_{1-x}$Ni$_{x}$)$_{2}$As$_{2}$. Note that the resistivity increase of the irradiated BaFe$_{2}$(As$_{1-x}$P$_{x}$)$_{2}$ crystals corresponds to $\delta R / R \sim 0.16$ [Ccm$^{-2}$]$^{-1}$, three times higher than for Ba(Fe$_{1-x}$Co$_{x}$)$_{2}$As$_{2}$.

\begin{figure}[t] \hspace{0.1cm}
\includegraphics[width=1.01\textwidth]{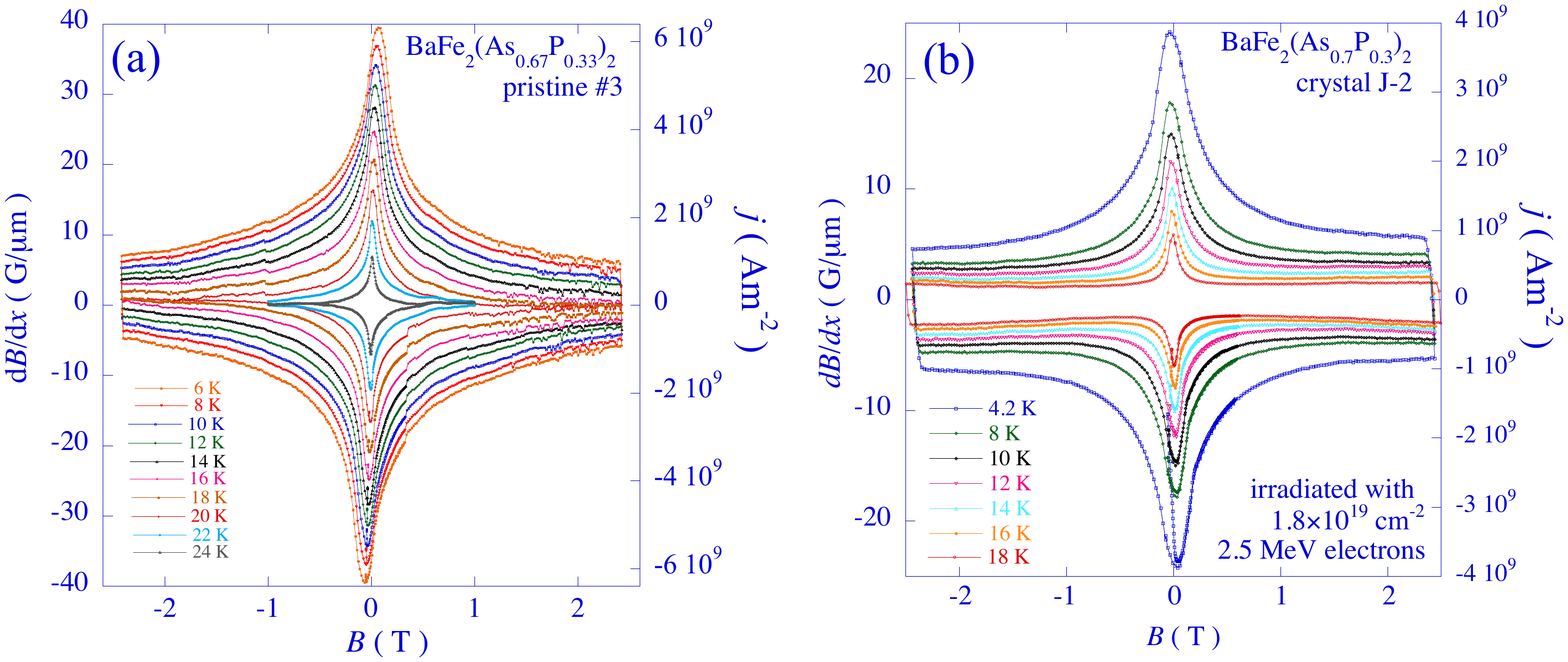}\vspace{-44mm}
\caption{\label{BaFeAsP-loops}(a) Hysteretic loops of the local flux density gradient versus local induction $B$, measured on the surface of a pristine BaFe$_{2}$(As$_{0.67}$P$_{0.33}$)$_{2}$ single crystal, at  various temperatures (indicated). The astroid-shaped hysteresis loops are determined by the sole strong-pinning contribution to the critical current. (b)  {\em ibid}, measured on the surface of a  BaFe$_{2}$(As$_{0.7}$P$_{0.3}$)$_{2}$ single crystal irradiated at 23~K with $1.8\times10^{19}$ electrons cm$^{-2}$, at the indicated temperatures. The opening of the loops at higher flux densities reveal the emergence of a weak collective pinning contribution by the atomic--sized point defects introduced by the irradiation.  }
\end{figure}

\begin{figure}[b]
\begin{minipage}{0.7\textwidth}\hspace{-0.2cm}
\includegraphics[width=1.25\textwidth]{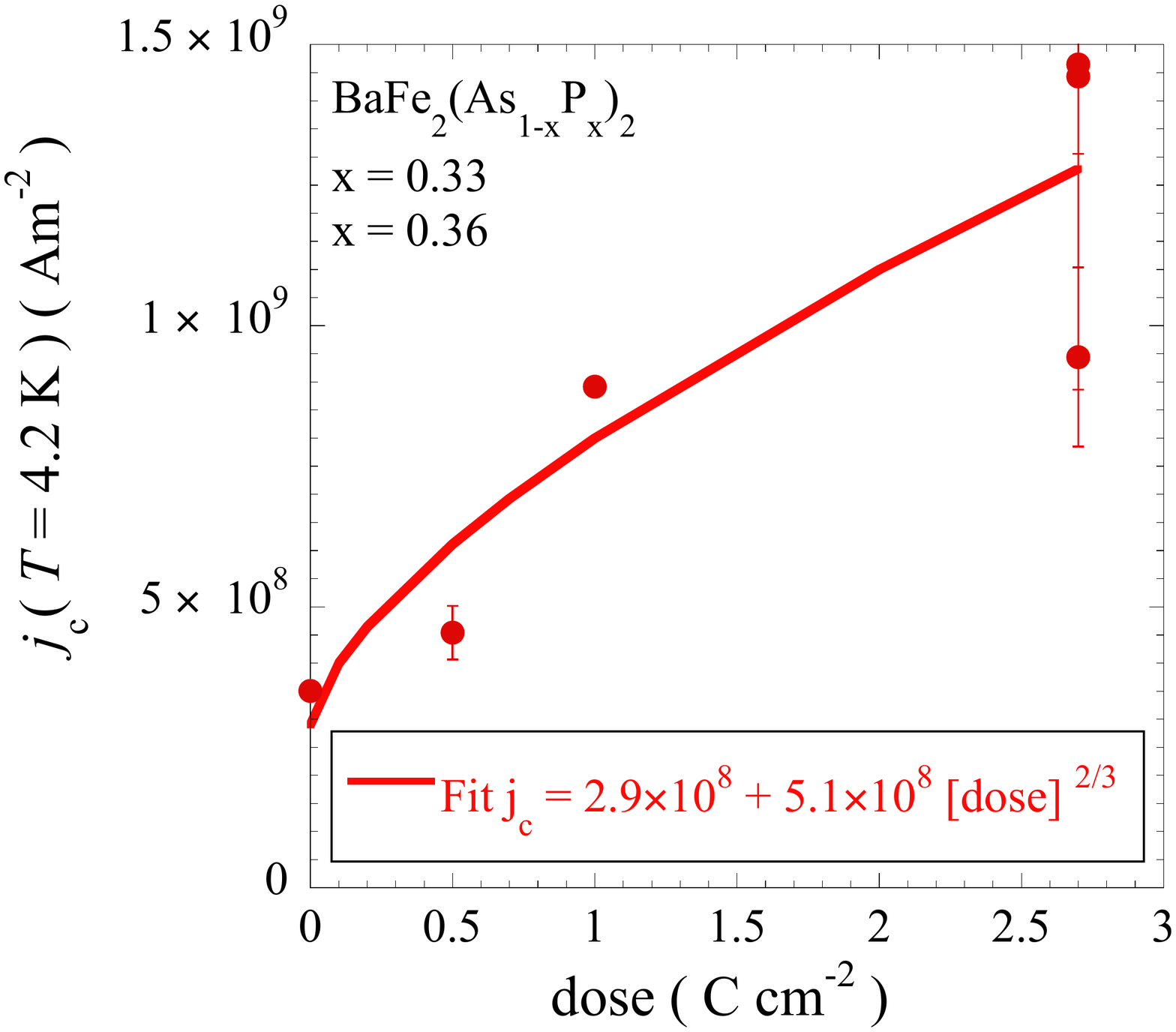}
\end{minipage}~\begin{minipage}[t]{9pc}\vspace{-45mm} 
\caption{\label{fig:jc-dose-BaFeAsP} BaFe$_{2}$(As$_{1-x}$P$_{x}$)$_{2}$ (with $x = 0.33$, $x = 0.36$): Electron-fluence dependence of the weak collective pinning contribution $j_{c}^{coll}$ to the critical current density. The drawn line denotes a fit to Eqs.~(\protect\ref{eq:jc-coll},\protect\ref{eq:Thuneberg}). Assuming that the point  defects most relevant for quasi-particle scattering are Fe vacancies induced by the irradiation, the parameter value $5.1\times 10^{8}$ would correspond to 0.0035 dpa / Ccm$^{-2}$. }
\end{minipage}
\end{figure}

Figure~\ref{BaFeAsP-loops} shows the hysteretic loops of the local flux density gradient on the surface of  a pristine (a) and electron--irradiated BaFe$_{2}$(As$_{1-x}$P$_{x}$)$_{2}$ single crystal (b), for different temperatures. The pristine single crystal shows astroid--shaped magnetic hysteresis  loops that are indicative of strong flux pinning by extended point-like defects only. After electron irradiation, the hysteresis loops open up due to the appearance of the field--independent collective pinning contribution from atomic-sized point defects.  We interpret the appearance of a non-zero $j_{c}^{coll}$ as being due to vortex pinning by vacancy-interstitial pairs introduced by the irradiation. Moreover, $j_{c}^{coll}$ monotonically increases (at low temperature) as function of irradiation fluence (Fig.~\ref{fig:jc-dose-BaFeAsP}), which allows for a direct comparison with the theory for quasi-particle scattering mediated collective vortex pinning. Under the assumption that the only unknown parameter, $\sigma_{tr}$, is again approximated by the ionic cross--section of  a Fe-vacancy, a fit to Eq.~(\ref{eq:jc-coll}) yields a defect density of 0.0035 dpa / Ccm$^{-2}$. This number is taken as more precise than that obtained for Ba(Fe$_{1-x}$Co$_{x}$)$_{2}$As$_{2}$ in subsection~\ref{section:Co}, in which only two points were available.  Note that the hypothesis of simple voids (non-scattering point defects) would necessitate an unphysical 1 dpa / Ccm$^{-2}$ to explain the magnitude of the critical current density change.

\begin{figure}[t] \hspace{0.1cm}
\includegraphics[width=1.01\textwidth]{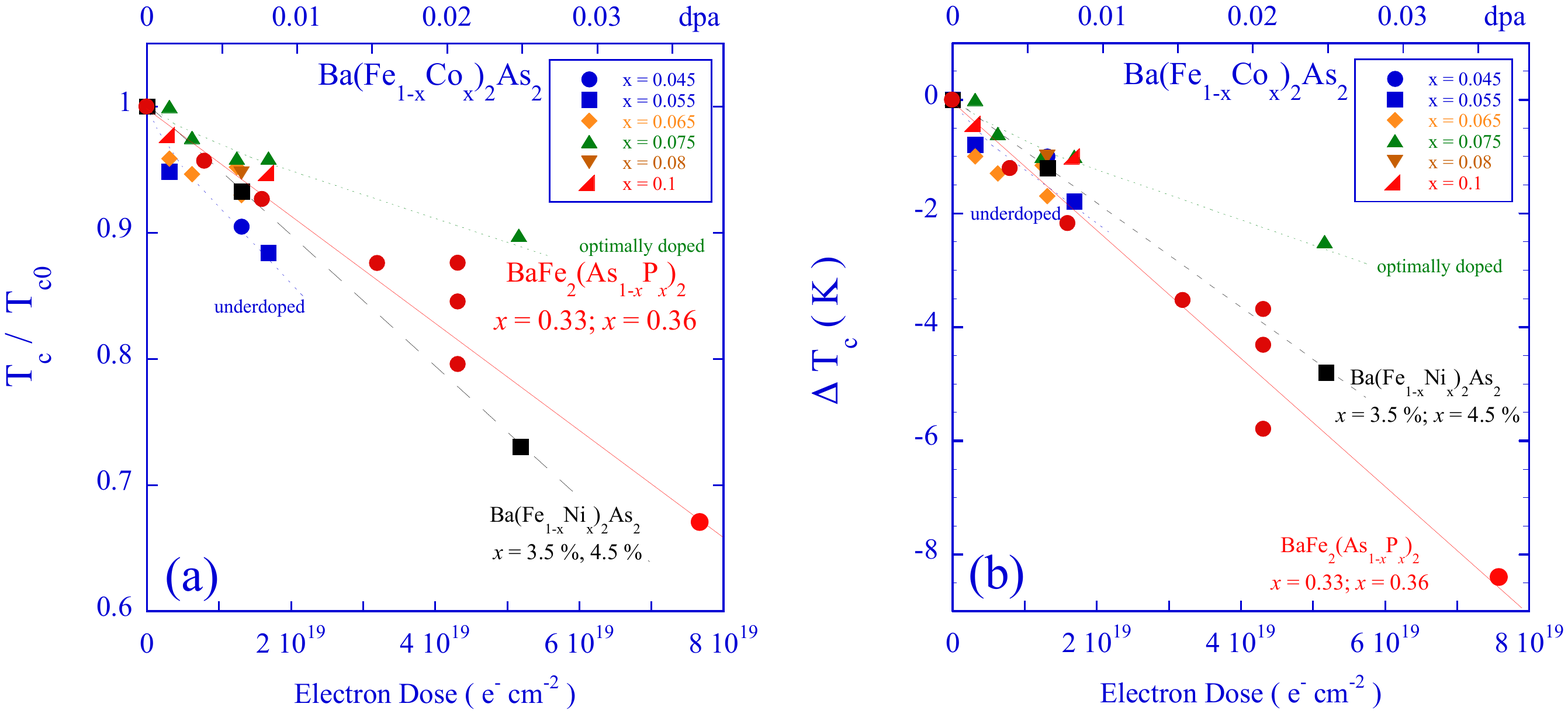}
\vspace{-49mm}
\caption{\label{Tc-vs-dose}(a)  Dose--dependence of the $T_{c}$--change of Ba(Fe$_{1-x}$Co$_{x}$)$_{2}$As$_{2}$, Ba(Fe$_{1-x}$Ni$_{x}$)$_{2}$As$_{2}$, and BaFe$_{2}$(As$_{1-x}$P$_{x}$)$_{2}$, after irradiation with 2.5 MeV electrons and annealing at 300~K. The upper scale shows the density of point defects (presumably Fe vacancies) added by the irradiation, such as determined from the dose-dependence of the critical current density of BaFe$_{2}$(As$_{1-x}$P$_{x}$)$_{2}$ (see Fig.\protect\ref{fig:jc-dose-BaFeAsP}). (b) Dose dependence of $T_{c}/T_{c0}$, where $T_{c0}$ is the critical temperature of the pristine crystal.  }
\end{figure}

\subsection{Discussion; Comparison between differently substituted materials}
The modification of the critical current density of Ba(Fe$_{1-x}$Co$_{x}$)$_{2}$As$_{2}$ and BaFe$_{2}$(As$_{1-x}$P$_{x}$)$_{2}$ by electron irradiation demonstrates that the field--independent contribution to $j_{c}$ apparent in sub-T to T--fields is indeed due to collective pinning by atomic-sized point defects. The very reasonable number obtained for defect generation by the electron beam,  0.35\% dpa / Ccm$^{-2}$, indicates that it is, in a first approximation, safe to assume that the most relevant produced defects are Fe vacancies, and that these vacancies are responsible for quasi-particle scattering. This interpretation is strengthened by the fact that scattering rates in Ba(Fe$_{1-x}$Co$_{x}$)$_{2}$As$_{2}$  estimated from the change of the resistivity are very much comparable to $\Gamma = n_{d}/[\pi N(0)] / \sin^{2} \delta_{0}$ estimated from the defect density. A comparison with Table~\ref{scattering}, shows that the scattering cross-section of Fe vacancies should be comparable to that of Co and Ru impurities.  Indeed, electron--irradiation of Ba(Fe$_{1-x}$Co$_{x}$)$_{2}$As$_{2}$  and Ba(Fe$_{1-x}$Ni$_{x}$)$_{2}$As$_{2}$  does not lead to appreciable changes of the temperature dependence of the superfluid density of those materials, which suggests that the effect of the dopant impurities overwhelms that of the defects added by the irradiation.

Irrespective of the precise changes in resistivity incurred by the irradiation, a striking feature of our results is the similar suppression of $T_{c}$ by the electron irradiation of underdoped Ba(Fe$_{1-x}$Co$_{x}$)$_{2}$As$_{2}$, Ba(Fe$_{1-x}$Ni$_{x}$)$_{2}$As$_{2}$ and BaFe$_{2}$(As$_{1-x}$P$_{x}$)$_{2}$ for a given dose. The $T_{c}$--suppression in optimally doped Ba(Fe$_{1-x}$Co$_{x}$)$_{2}$As$_{2}$ is a factor 2 less. Given the clear evidence for line nodes in BaFe$_{2}$(As$_{1-x}$P$_{x}$)$_{2}$, this poses the question possible nodal structures of the order parameter in Ba(Fe$_{1-x}$Ni$_{x}$)$_{2}$As$_{2}$, and, possibly, Ba(Fe$_{1-x}$Co$_{x}$)$_{2}$As$_{2}$ as well. In the latter two materials, it was proposed that a nodal line might exist on the $\alpha$--(hole--like) sheet, at finite $k_{z}$ \cite{Reid2010,Martin2010}.

It is tempting to compare our results to that obtained by other types of irradiation, as well as by doping. It turns out that the suppression of $T_{c}$ for optimally doped Ba(Fe$_{1-x}$Co$_{x}$)$_{2}$As$_{2}$ is remarkably comparable to that reported for Co-doping of Ba$_{1-x}$K$_{x}$Fe$_{2}$As$_{2}$ in Ref.~\cite{Wang2012}, and much less than that obtained by Co-doping of KFe$_{2}$As$_{2}$ studied by the same authors. The decrease of $T_{c}$ observed after electron irradiation is also comparable, as far as optimally doped Ba(Fe$_{1-x}$Co$_{x}$)$_{2}$As$_{2}$ is concerned, to the Co-doping experiments on Ba$_{1-x}$K$_{x}$Fe$_{2}$As$_{2}$ by Li {\em et al.}, while the $T_{c}$ suppression in underdoped Ba(Fe$_{1-x}$Co$_{x}$)$_{2}$As$_{2}$, Ba(Fe$_{1-x}$Ni$_{x}$)$_{2}$As$_{2}$, and  BaFe$_{2}$(As$_{1-x}$P$_{x}$)$_{2}$ is similar to the Cu and Zn doping by those authors. Even if the comparison with chemical doping is tenuous, the observed trend clearly indicates a much weaker sensitivity of the materials studied here than what is expected for the scenario of $s_{\pm}$ superconductivity with strong interband scattering, and sets  the three materials even further from the $d$-wave scenario.

A comparison with the previous irradiation studies reveals that  $T_{c}$--suppression after electron irradiation is somewhat weaker than the results obtained by Nakajima by 3 MeV proton irradiation of optimally doped Ba(Fe$_{1-x}$Co$_{x}$)$_{2}$As$_{2}$ \cite{Nakajima2010}, and much weaker than in the $\alpha$--particle irradiation of NdFeAs(O,F) by Tarantini {\em et al}. \cite{Tarantini2010}.

\section{Summary and Conclusions}

Ba(Fe$_{1-x}$Co$_{x}$)$_{2}$As$_{2}$ crystals of different dop"ng levels $x$, as well as Ba(Fe$_{1-x}$Ni$_{x}$)$_{2}$As$_{2}$ and BaFe$_{2}$(As$_{1-x}$P$_{x}$)$_{2}$ crystals, have been subjected to high energy (2.5 MeV) electron irradiation at low temperature (21 K). In spite of substantial annealing and clustering of the defects, the enhancement of the collective pinning contribution to the critical current density due to quasi-particle scattering in the vortex cores in Ba(Fe$_{1-x}$Co$_{x}$)$_{2}$As$_{2}$, and its appearance in BaFe$_{2}$(As$_{1-x}$P$_{x}$)$_{2}$ in which it is absent in the pristine material, demonstrate that the irradiation produces atomic-sized point defects. The increase in magnitude of the low-temperature critical current density is consistent with the pinning by Fe vacancies, created by the irradiation. In Ba(Fe$_{1-x}$Co$_{x}$)$_{2}$As$_{2}$ and Ba(Fe$_{1-x}$Ni$_{x}$)$_{2}$As$_{2}$, the introduction of these supplementary point defects does not lead to a significant modification of the temperature dependence of the superfluid density, which suggests that the defects are overwhelmed by pre-existing disorder (notably, the presence of the dopant atoms). Surprisingly, the critical temperature is similarly suppressed in all three materials. This, in spite of the fact that the order parameter in BaFe$_{2}$(As$_{1-x}$P$_{x}$)$_{2}$ is thought to have line nodes, while this possibility is much less certain in the other two materials. The results therefore lend credence to evidence for line nodes obtained from $c$-axis penetration depth \cite{Martin2010} and thermal conductivity measurements \cite{Reid2010}.

\section*{Acknowledgements} 

We acknowledge support from the the grant ÒMagCorPnicÓ of the R\'{e}seau Th\'{e}matique de Recherche Avanc\'{e}e ``Triangle de la Physique du Plateau de Saclay'', and by the Agence Nationale de la Recherche grant ``PNICTIDES''. Part of the work was also possible thanks to the support of the ECOS-Sud-MINCyt France-Argentina bilateral program, Grant No.A09E03. Work in Japan was supported by KAKENHI from JSPS, and by Grant-in-Aid for the Global COE program ÔÔThe Next Generation of Physics, Spun from Universality and EmergenceÕÕ from MEXT, Japan.

\end{document}